\documentclass{llncs}

\usepackage{epsfig}
\usepackage{algorithm}
\usepackage[noend]{algorithmic}

\usepackage{algorithm}
\usepackage[noend]{algorithmic}
\usepackage{graphicx}
\usepackage{subfigure}
\usepackage{amssymb}
\usepackage{amsmath}
\usepackage{stmaryrd}

\newtheorem{ru}{Rule}

\begin{document}

\title{Simulating Opinion Dynamics in Heterogeneous Communication Environments}

\author{Walter Quattrociocchi\inst{1}\inst{2} \and Rosaria Conte\inst{2} \and Elena Lodi\inst{1}}

\institute{University of Siena - Department of Mathematics and Computer Science ``R.Magari'' - Siena, Italy
\and Laboratory of Agent Based Social Simulation - Institute of Cognitive Sciences and Technologies - CNR Rome, Italy}

\maketitle

\begin{abstract}
Since the information available is fundamental for our perceptions and opinions, we are interested in understanding the conditions allowing for a good information to be disseminated.
This paper explores opinion dynamics by means of multi-agent based simulations when agents get informed by different sources of information. The scenario implemented includes three main streams of information acquisition, differing in both the contents and the perceived reliability of the messages spread. Agents' internal opinion is updated either by accessing one of the information sources, namely media and experts, or by exchanging information with one another. They are also endowed with cognitive mechanisms to accept, reject or partially consider the acquired information.
We expect that peer-to--peer communication and reliable information sources are able both to reduce biased perceptions and to inhibit information cheating, possibly performed by the media as stated by the agenda-setting theory. In the paper, after having shortly presented both the hypotheses and the model, the simulation design will be specified and results will be discussed with respect to the hypotheses. Some considerations and ideas for future studies will conclude the paper.
\end{abstract}

\section{Introduction}
What is the effect of different information sources and communication processes on agents' opinions? Since the early thirties (\cite{Sherif88,Asch55}) the informal study of social influence has produced abundant evidence of structural factors affecting people's beliefs. In social life, agents are exposed to different communication systems interacting with one another, ranging from one-to-many information transmission typical of traditional broadcasting media, to one-to-one systems characterizing the new media, but including several intermediate modalities. How do they interact? Which one is most likely to exercise the strongest influence on agents' opinions? Despite the importance social scientists attribute to the role of persuasive communication (think of the Hovland school of persuasion), different communication systems have rarely been compared under natural conditions, and even less in artificial experiments. Based on social impact theory (\cite{nowak90}) recent simulation-based studies of opinion dynamics \cite{Staufer2007,Lorenz07,niguez09opinion,Castellano2007,Hu09,brunetti2010,AQ2010a,QuattrociocchiPC09} observe how numerically defined opinions spread and aggregate over a given population as a function of the distance among the values agents assign to them. Within these studies, however, the process of communication among the agents is not explicitly addressed. Plunged into the same network, agents are assumed to exchange opinions as a function of the distance between them: the lower this is, the more the agents are inclined to converge. 

In this paper, the role of different forms of communication in opinion dynamics is addressed with the help of agent based simulation. Opinions are numerically defined on one parameter that stands for the certainty with which agents hold them; as in the bounded confidence model (\cite{amblard01}), agents are assumed to exchange their opinions based on the distance between them. However, we introduced three modifications over the preceding works on opinion dynamics: agents (a) share two relatively independent opinions, (b) are exposed to different forms of communication, namely one-to-many and one-to-one, which will be characterized later on in the paper on a number of dimensions,  and (c) receive inputs from two distinct sources of information, expert and non-expert. 

The first modification is suggested by the necessity to make the scenario more realistic. 

The latter modifications instead are needed to address a topical question. It is a common opinion nowadays that the new media play a positive role in the control and improvement of the quality of information circulating in a system. To what extent is this opinion backed up by existing evidence? Comparing the effects of peer-to-peer communication Vs traditional media on the diffusion of opinions and observing their interaction, we aim to investigate whether the former system is bound to amplify the effects of the latter, or can exercise a relatively independent influence on information quality. To clearly distinguish the effects of the two systems, we have introduced the distinction between the expert and the non-expert source. To what extent will the experts affect existing opinions? How much expert-driven information must be accessed through the new media for the average quality of information circulating in the population to increase?

\section{Within Different Communication Paradigms}
Information is acquired both through communication among agents and from the centralized media. One-to-one and one-to-many communication have been compared by marketing scientists (\cite{Deighton95}) as well as cultural evolutionary scientists (\cite{cavalli-sforza86}), and shown to have different although balanced effects.
As reported in \cite{eRep}, according to classic communication models (see \cite{Hoadley99}), peer to peer (P2P) communication is a step-wise, asynchronous process that requires a more or less lengthy temporal extension: it starts at one point in time and takes effect after a certain amount of steps, and in each of them it is iterated. On the contrary, centralized or broadcast communication takes place at once. Whereas broadcast communication is a one-to-many process (\cite{Harris04}), P2P communication reaches a smaller audience.
As summarized in table \ref{tab:pvsbc}, two complementary patterns of properties emerge. Each pattern allows different expected performances. P2P systems are less efficient and more liable to corruption, although more interactive and controllable. P2P can either be proactive or reactive, whilst broadcast (BC) communication generally is only proactive. The former is based upon and aimed at reciprocating information, whereas broadcast communication can hardly be expected to be reciprocated (it is often institutional). P2P is spontaneous and based upon acquaintanceship or familiarity networks, while BC communication, which is based upon other types of networks, is often facilitated or allowed by the sharing of new technologies.

\begin{center}
\begin{table}
\centering
\label{tab:pvsbc}
 \begin{tabular}{|c|c|}
\hline 
Peer-to-Peer Communication (P2P) & Broadcast Communication (BC)\tabularnewline
\hline
\hline 
one to one/few & one to many\tabularnewline
\hline 
asynchronous & synchronous\tabularnewline
\hline 
step wise & at once\tabularnewline
\hline 
sensitive to temporal extention  & not affected by temporal extention\tabularnewline
\hline 
interactive & proactive\tabularnewline
\hline 
based upon reciprocation & no reciprocation\tabularnewline
\hline 
based upon familiarity networks & not affected by the network topology\tabularnewline
\hline 
\end{tabular}
\caption{A comparison between Peer to Peer and Broadcast communication properties}
\end{table}
\end{center}

Which further consequences can we expect from either system? How deep is their respective influence on the population? How do they interact, when insisting on the same population? Are they interdependent, or is there a dominance of one system on the other, and if so, which one is more influential? Finally, what is the effect of their interaction on information quality?

\section{Related Works}
The effect of communication on opinion formation has been addressed by different disciplines from within the social and the computational sciences, as well as complex systems science. 
Social scientists focus on polarization, i.e. the concentration of opinions by means of interaction, as one main effect of the "social influence'' \cite{festinger50}. 
Social psychology offers an extensive literature on attitude change models, as reviewed by \cite{Mason07}. 
Most influential in social psychology is the ``The Social Impact Theory'' \cite{Latane90}, according to which the amount of influence depends on the distance, number, and strength (i.e., persuasiveness) of influence sources. As stated in (\cite{Castellano2007}), an important variable, poorly controlled in current studies, is structure topology. Interactions are invariably assumed as either all-to-all or based on a spatial regular location (lattice), while more realistic scenarios are ignored.

The most popular model applied to the aggregation of opinions is the bounded confidence model, presented in \cite{amblard01}.

Much like previous studies, in this paper agents exchanging information are modeled as likely to adjust their opinions only if the preceding and the received information are close enough to each other. Such aspect is modeled by introducing a real number $\epsilon$, which stands for tolerance or uncertainty (\cite{Castellano2007}) such that an agent with opinion $x$ interacts only with agents whose opinions is in the interval $] x - \epsilon ,  x + \epsilon[$.

The model we present in this paper extends the bounded confidence model by providing agents with two, instead of one, conflicting and independent values representing their opinions about, say, welfare and security. 
Furthermore, in the model agents resort to two additional sources of information, external to the social network, aimed at representing experts and media.

\subsection{Our Previous Works}

In our previous works \cite{quattrociocchi2008}, \cite{quattrociocchi2008b}, \cite{quattrociocchi2009a,quattrociocchi2009b}, \cite{Balke08} and \cite{eRep} we investigated the role of communication systems on agents' perceptions, by means of multi-agent based simulations, when informational cheating occurs.

In particular in \cite{quattrociocchi2010d} we focused on the impact of peer-to-peer communication on the quality of information when central media spread false information. 
Our simulations show that the wider the audience reached by the broadcasting system, the stronger its influence, especially when people are poorly self-confident and more likely to accept incoming information. We also showed that for reasonable levels of confidence, P2P communication inhibits the effects of media broadcasting until this has reached the 40\% of the audience. 
Since the media plays a fundamental role in the opinion formation of agents, in this paper we extend the model by introducing an additional  source of information differing from media for both the kind and the perceived reliability of the delivered messages.

\section{Research Questions}

In our previous work \cite{quattrociocchi2010d} the focus was on the interplay between institutional broadcasting and P2P communication. 

In particular, in our simulation we model the correlation within the frequency of information delivered by the media and their consequent effects on social perceptions. This phenomenon, called agenda-setting, has been theorized in 1972 by Maxwell McCombs and Donald Shaw in \cite{Maxwell}.
In that model, agents were exposed to both the conventional media, repeating the same message at each time step, and the new media, i.e. the information circulating in the neighborhood. 

The results obtained, pointing out the role of both communication systems on the aggregation of opinions, encouraged us to model another source of information, the expert. Agents sometimes have direct access to experts, for example via Internet. In this paper, we will assume that, when accessing Internet, the agents are quite confident in the truth-value of the acquired information. Of course, this assumption is somewhat arbitrary, but what matters here is the source of influence rather than the way it is found out.

Hence, the follow-up research questions addressed within the present work: (a) what would happen to agents' opinions if both conventional and new media were confronted with an additional P2P-based source of information highly trusted by the agents? (b) Which is the role of the white-zone, namely the percentage of agents that are reached neither by media nor on the Internet? Furthermore, how many experts are needed in a network to reduce the information asymmetry between agents and conventional media, or, to put it more explicitly,  how many experts are needed to contrast possible informational cheating spread within the system? (d) How does the interaction topology, i.e. the network structure, affect the information diffusion and the dynamics of opinions? In this paper we address the former three sets of questions, leaving the last one for future works.

\section{Definitions}

In the simulated system, traditional media send out messages in broadcasting to a variable percentage of the population, while members communicate with neighbors and are thus exposed to an additional source of information which they consider highly reliable, the experts.

In this section the basic definition related with our model and the simulation results are provided.

\subsection{Scale Free Network}

Let $N$ be a connected graph, namely a scale free network, in which each node $v$ has a number of $k$ originating links following a power law distribution $P(k)\sim k^{-y}$.
We generate a scale free network by progressively adding nodes on a previously existing network and then introducing links to the existing nodes following the so called ``preferential attachment'' mechanism. The construction strategy of the algorithm aims at maintaining the link probability between any couple of nodes proportional to the number of existing links $k_{i}$ already connected to the selected node.


\subsection{The General Bounded Confidence}

As mentioned above, the most famous model of opinion dynamics based upon bounded confidence is the one developed by Deffuant and Weisbuch in \cite{amblard01}. The model can be explained as an asynchronous game in a distributed environment where the nodes $v$ of the network $N$, i.e., the agents, interact by exchanging their opinions.
For instance, consider the iterative process over $N$, where each element $v \in N$ updates its internal state at each time step by comparing its opinion with the information circulating in the neighborhood. From a theoretical point of view, this model presents a number of oversimplifications. In particular, it assumes that low distance is the only determinant of opinion update. Other aspects, for example, (a) the degree of certainty on one's opinion and (b) the extent to which it is shared by others are ignored. The second aspect will be addressed in future works. For the time being, we limit ourselves to extend the original model to a slightly more complex situation in which agents have a generic subjective disposition to accept others' beliefs, and hold two independent opinions. Hence, given two agents $x$ and $y$ exchanging their opinions $v(x)$ and $v(y)$ the entities' internal states are updated by applying the following rule:

\begin{ru}
\label{rule1}
\begin{center}
\begin{equation}
\begin{split}
if \mid v(x)-v(y)| \leq t\\
 v(x) = v(x) + m (v(y)-v(x))    
\end{split}
\end{equation}\end{center}
\end{ru}

Let $m$ be a fixed value within the interval $(0..0.5]$ representing the convergence parameter, i.e. a way to increase the convergence by dividing the distance $d = v(x)-v(y)$ between two agents' opinions in $\frac{1}{m}$ steps. 

The variable $t$ represents agents' tolerance, a threshold defining the limit under which an opinion can be accepted by an agent.

\section{Extending The Bounded Confidence Model}
\begin{figure}[h!]
 \centering
 \includegraphics[width=210mm,bb=0 0 1244 392]{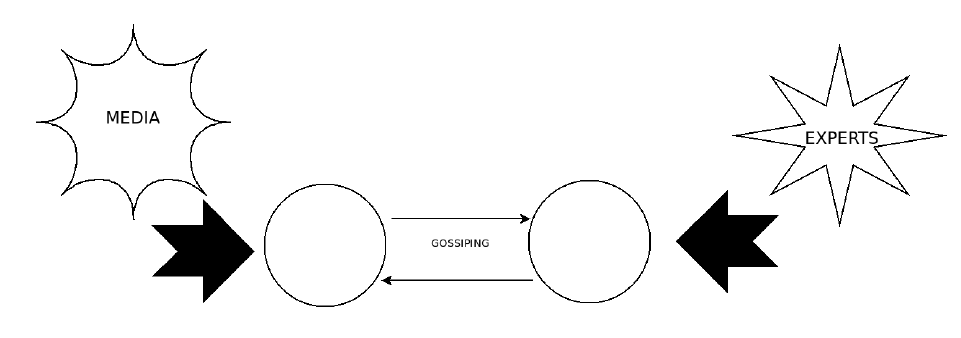}
 \caption{The Communication Model}
 \label{fig:interaction}
\end{figure}

Our model extends the model of Deffuant by providing agents with two conflicting values representing their opinions: one related to welfare and one related to security. Furthermore, as shown in \textbf{Figure \ref{fig:interaction}} in the model agents get informed by accessing two different additional sources of information: experts and media.

We refer to agents accessing the former source as wise agents, while agents accessing the latter are called televiewers. 

In general, to process information, agents apply the rule of bounded confidence; on the contrary, a wise agent is more inclined to accept the information acquired by the experts.

\subsection{Entities of the Model}
In this section the agents protocol and their interaction patterns will be introduced.

\subsubsection{Media.}
Conventional media is simulated as a special agency, reporting the same message at each simulation turn to a subset of the agents' set $V$  .  

The media agent $m\in M$ is not linked to the network and has the goal to persuade the audience that security is a matter more important than welfare. The media's reported message is denoted by the following set:

\begin{equation}
\begin{split}\left\lbrace m_{l},m_{r},V_{1}\mid0\leq m_{l}\leq1\wedge0\leq m_{r}\leq1,V_{1}\subset V\right\rbrace \end{split}
\end{equation}

where $m_{l},m_{r}$ represent the media reported values of events respectively related to welfare and security issues.

\subsubsection{Interacting Peers.}
The audience is composed by agents (i.e. the network nodes), whose shared goal is to exchange information with other agents in one's neighborhood. They also receive messages from the media and from the wise agents.

When interacting with one another, agents  $v\in V$ are provided with an internal state defined as follows:

\begin{equation}
\left\lbrace v_{l},v_{r}\mid0\leq v_{l}\leq1\wedge0\leq v_{r}\leq1\right\rbrace 
\end{equation}

respectively representing agents' beliefs about welfare and security issues. 
The closer $v_{l}$ and $v_{r}$ values are to 1, the more each debated issue (i.e. welfare and security) is considered to be important by the agent considered.

The agents' internal state, in the protocol, corresponds to the message sent as an answer to each external request of information coming from its neighbours. 

The initial configuration of agents' opinions is set up according to a uniform random distribution.

\subsubsection{Wiseagents and Televiewers.}

The set of agents $V$ is composed by two kinds of agents, each subpopulation is denoted by the source of information accessed.

Agents accessing experts will be called WiseAgents (WAs), those exposed to media will be called TeleViewers (TVs). 

WiseAgents and TeleViewers differ in the way they process the information acquired. The former are highly confident in its truth value while the latter process the media information as the information they receive from peers, i.e. by applying the bounded confidence model mechanism.

\subsection{The Model Interaction}

\subsubsection{Media and Peers.}

At each simulation turn the media agency sends the televiewers the same message $\left\lbrace e_{l},e_{r},V_{1}\right\rbrace $. 

Agents acquire information from media agencies according to a passive protocol, by acquiring the values they send and comparing them with their previous opinions. 
Information is either accepted or not, based on the bounded confidence mechanism (\cite{amblard02}). 

The agent's opinions $v_{l},v_{r}$ and the information from the media $m_{ld},m_{rd}$ are transformed in two new agent's opinions. The function generates two new values for $v_{l},v_{r}$.
The variable $t\in\Re$ stands for peer agents' tolerance, i.e., the subjective disposition to accept others' information. The two
guard variables $g_{l}$ and $g_{r}$ are calculated by the Boolean expression returning true if the difference between
acquired and owned information is below the tolerance threshold $t$.
The guard variables $g_{l}$ and $g_{r}$ respectively control the access to the updated values of $v_{l}$ and $v_{r}$, which is implemented on two independent opinion spaces.
The values of $v_{l}$ and $v_{r}$ are updated through the following:

\begin{center}
\[
\begin{split}v_{l}=(v_{l}+(m_{ld}(v_{l}-m_{ld}))\\
v_{r}=(v_{r}+m_{rd}(v_{r}-m_{rd}))\end{split}
\]
\par\end{center}

\subsubsection{Among Interacting Peers.}

Agents exchange information by comparing their preferences. This interaction is executed after both TeleViewers and WiseAgents receive the information by their respective sources.

Each agent communicates with the set of neighbours within a distance set to 1. We will follow  \cite{amblard02}' convention, according to which, when communication occurs between any two agents, these mix their opinions when the differences is smaller than the threshold $t$. For example, assuming that two agents' opinions are respectively represented by $x$ and $x_{i}$ if $\mid x-x_{i}\mid\leq t$ their opinions will be mixed by applying the \ref{rule1} rule on both the agents preferences.

\section{Experiments}

The experiments design has been performed with the main aim to perform a what-if analysis, based upon simulations, on the effect of information and communication on social perceptions.

We focus on the effect of two different sources of information, reporting different (and complementary) messages to the audience, on the agents' opinions trend. In particular we stress the polarization of opinions toward one of the two main debated issues, which in the model corresponds to security and welfare.

The main goal of the media is to persuade the audience that security is more important than welfare (reporting the same message respectively fixed to 8 for security and 3 for welfare). On the contrary, agents resorting to experts (giving respectively 3 for security and 8 for welfare) consider welfare more desirable than security.

In addition, experts and media differ as to the way their information is trusted by the agents. On the one hand, Wise Agents accept the information provided by the experts without applying the proviso of the bounded confidence model, meaning that the information is assumed as truthful. 
On the other hand, TeleViewers, when acquiring information from the media, adjust their beliefs according to the bounded confidence model.

Below, we will discuss the initial results of our model, in order to check whether it is of any utility in addressing the research questions raised within this work.

\subsection{Scenario 1: Media and Experts}

The first set of scenarios aims at exploring the consequences of different percentages of agents exposed to the main streams of information. 
Hence, each scenario is characterized by an increasing degree of exposure to conventional media and by a decreasing number of experts.

As one can see from the experimental settings listed in details in Table \ref{tab:expII}, the population is composed by 100 agents (NA) endowed with the ability to process the information through a variable level of confidence (TOL). A subset of agents (WAs) can access a different source of information with values different from the ones reported on by the media (TVs).
Each scenario has been simulated in ten runs.
In short these experiments are aimed at understanding the mutual effects of different information (with complementary values) delivered by different agencies.

\begin{table}[h!]
\centering
\caption{Experiments Settings }
\begin{tabular}{|c|c|c|l|} \hline
NA&MB&WAs&TOL\\ \hline
100 & 0& 100&0.2,0.5 and 0.8\\ \hline
100 & 10& 90&0.2,0.5 and 0.8\\ \hline
100 & 20& 80&0.2,0.5 and 0.8\\ \hline
100 & 30& 70&0.2,0.5 and 0.8\\ \hline
100 & 40& 60&0.2,0.5 and 0.8\\ \hline
100 & 50& 50&0.2,0.5 and 0.8\\ \hline
100 & 60& 40&0.2,0.5 and 0.8\\ \hline
100 & 70& 30&0.2,0.5 and 0.8\\ \hline
100 & 80& 20&0.2,0.5 and 0.8\\ \hline
100 & 90& 10&0.2,0.5 and 0.8\\ \hline
100 & 100& 0&0.2,0.5 and 0.8\\ \hline
\end{tabular}
\label{tab:expII}
\end{table}

\subsection{Emerging Results from Media and Experts}

\begin{figure}[h!]
 \centering
 \subfigure[Tolerance: 0.2]
   {\includegraphics[width=40mm]{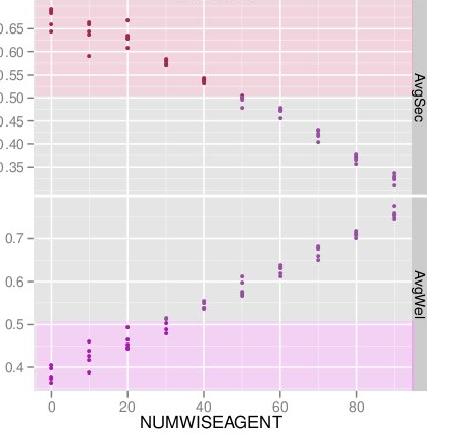} \label{fig:IIWISE02}} 
 \hspace{1mm}
 \subfigure[Tolerance: 0.5]
  {\includegraphics[width=40mm]{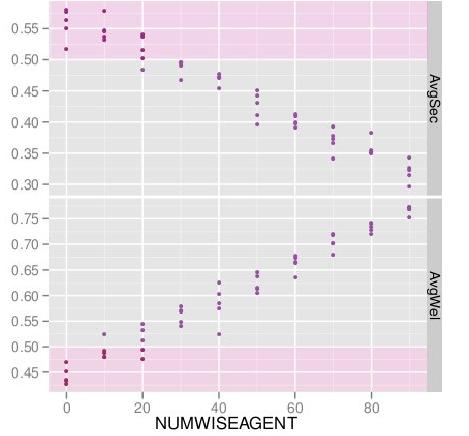} \label{fig:IIWISE05}} 
\hspace{1mm}
\subfigure[Tolerance: 0.8]
  {\includegraphics[width=50mm]{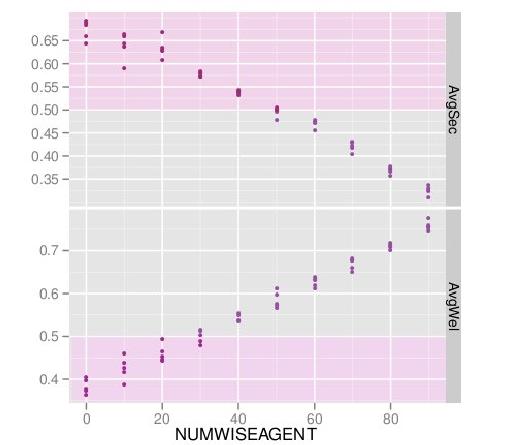} \label{fig:IIWISE08}} 
 \caption{Opinion Trend under Peer Pressure without Black Zone}
\label{fig:expIIP}
\end{figure}

The results emerging from the first battery of experiments are shown in \textbf{Figure \ref{fig:expIIP}} where the aggregate values of both agents' opinions on security and welfare matters, for increasing presence of WAs, are reported.
The figures report the opinion trends for different levels of tolerance (0.2, 0.5, 0.8). 

All figures show the same effect: the higher the presence of WAs the more the agents' opinions converge toward the values reported on by the experts (3 for security and 8 for welfare). On the contrary, namely when the media reach the largest amount of the total population (and WAs are on the lowest level), no convergence can be observed but only the reduction of the distances among values. 
In our view this phenomenon indicates that, consistent with our previous results, the efficacy of traditional media are sensitive to the effect of peer-to-peer communication. By definition WAs do not process the information with the bounded confidence mechanism, thus the tolerance does not affect opinion adjustment.

Notice the extent to which the opinion trend passes the line of the average value (i.e. $0.5$). The WAs effect is relevant when the global tolerance is fixed to 2: the 30\% of WAs is sufficient to invert the effect on mass opinions of the 70\% of TWs.
Furthermore when tolerance is higher (5), the number of WAs needed to inhibit media is smaller (20\%). 
One may expect that tolerance plays a fundamental role in opinions' convergence toward the media or the WAs values.
Results indicate that the high level of confidence in buying others' information has a side effect: WAs are less efficient when tolerance is high, meaning that if uncertainty is strong agents are inclined to accept any information as truthful, even that provided by the media.

\subsection{Scenario 2: Media, Internet and Uninformed Agents}

The second scenario is similar to the previous one, except for the diverse distribution of WAs and TVs. 
In the preceding experiment the total amount of the population is either reached by the media, or resorts to experts. 
Peer-to-peer communication inhibits the spreading of information from the media, but what would happen if a given percentage of agents is reached neither by the media nor by the experts? In other words, let us assume that in the scenarios characterized by an increase of TVs (and consequently by a decrease of WAs), a 30\% of the agents, that we call the white zone, use only the information circulating in the neighborhood as their main source of information.

As one can see from the experimental settings listed in details in Table \ref{tab:expIII}, the population is composed by 100 agents (NA) endowed with the ability to process the information through a variable level of confidence (TOL). A subset of agents (WAs) can access a different source of information with values different from the ones reported on by the media (TVs).
This battery of experiments aims at understanding the effect of the two sources of information also on agents that are never directly reached by them. 
In particular the agents lying in the white zone will receive the information delivered either by experts and media but only through their neighbors. Such a process of information transmission propagates information through all the interaction topology: a message delivered by an agent will get far away on the network. 

Each scenario has been simulated in ten runs.
\begin{table}[h!]
\centering
\caption{Experiments Settings }
\begin{tabular}{|c|c|c|l|} \hline
NA&MB&WAs&T\\ \hline
100 & 0& 70&0.2,0.5 and 0.8\\ \hline
100 & 10& 60&0.2,0.5 and 0.8\\ \hline
100 & 20& 50&0.2,0.5 and 0.8\\ \hline
100 & 30& 40&0.2,0.5 and 0.8\\ \hline
100 & 40& 30&0.2,0.5 and 0.8\\ \hline
100 & 50& 20&0.2,0.5 and 0.8\\ \hline
100 & 60& 10&0.2,0.5 and 0.8\\ \hline
100 & 70& 0&0.2,0.5 and 0.8\\ \hline
\end{tabular}
\label{tab:expIII}
\end{table}

\subsection{Emerging Results from Media, Internet and Uninformed Agents}
\begin{figure}[h!]
 \centering
 \subfigure[Tolerance: 0.2]
   {\includegraphics[width=40mm]{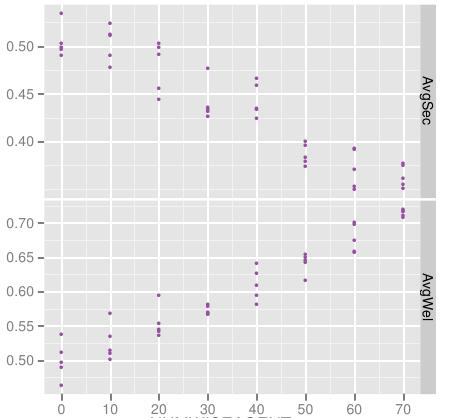} \label{fig:IIIWISE02}} 
 \hspace{1mm}
 \subfigure[Tolerance: 0.5]
  {\includegraphics[width=40mm]{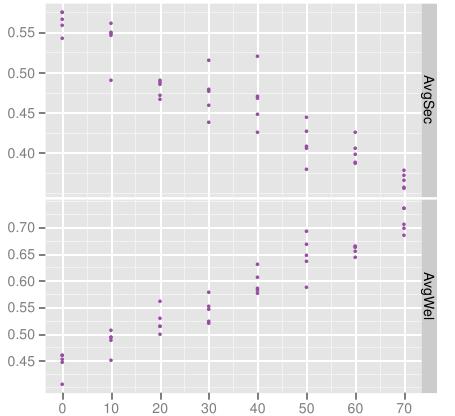} \label{fig:IIIWISE05}} 
\hspace{1mm}
\subfigure[Tolerance: 0.8]
  {\includegraphics[width=40mm]{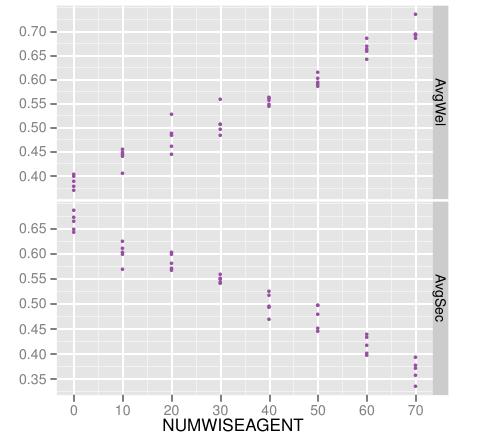} \label{fig:IIIWISE08}} 
 \caption{Opinion Trends with the 30\% of agents being not reached by media and experts}
\label{fig:expIIIP}
\end{figure}

The results emerging from this set of experiments are shown in \textbf{Figure \ref{fig:expIIIP}}, where the aggregate values of agents' opinions with respect to welfare and security are reported for the various different scenarios, each one denoted by an increasing number of WAs and a decreasing number of TVs.
The figures reported on the different scenarios stand for different values of tolerance (0.2, 0.5, 0.8). Remember that tolerance is the threshold representing the limit within which an information can be taken into account by the agents or not. Hence the higher the tolerance  the more likely the agents will be to trust informers.
As derived from our hypotheses, the white zone matters. 
Looking at the initial portion of each box in \textbf{Figure \ref{fig:expIIIP}}, in which all agents are TVs (and WAs are absents), the average values for security and welfare are never the same as the messages spread by the media. 

There is a clear evidence of the capacity of peer-to-peer communication to inhibit the effect played by the central media. Furthermore, this evidence is amplified by the white zone (i.e. the 30\% of agents not directly reached by the media nor by the experts). 
Looking at the pictures shown in \textbf{Figure \ref{fig:expIIP}}, the distance between the initial values of opinions and the media's message is smaller than the same distance in the second scenario. Looking at the pictures reproduced in \textbf{Figure \ref{fig:expIIP}}, the distance between the initial values of opinions and the media's message is smaller than the same distance in the second scenario.
Even when TVs are predominant, the central media do not lead to opinions converging on the values they transmit.

\section{Conclusions}
In this paper, the dynamics of two relatively independent opinions in a simulated network is observed. Plunged into the network, agents characterized as more or less likely to exchange opinions with neighbors are also exposed to information broadcasted by central media. 

A first series of experiments shows that when central media spread false news, P2P communication can reduce the effect of informational cheating until the broadcasting message has reached around half the population, but it tends to lose this compensating effect for increasing values of agents' exposure to informational cheating. Even a small number of experts can dramatically re-orient agents' opinions. This effect is less flashy when agents are more likely to accept others' opinions, what should not come as a surprise: the less confident agents are, the more they tend to oscillate among different opinions. Instead, the more confident they are, the lesser they are likely to converge on either the experts' or anyone else opinions. However, with a mild level of confidence the expert source is more efficacious in contrasting the impact of informational cheating.

A second series of experiments shows the effect of peer-to-peer communication. When a certain percentage of the population is not directly reached neither by the media nor by the experts, the agents' opinions do not totally converge on the messages spread by the media, not even for the highest number of agents exposed to the media and the lowest number of agents' accessing the experts. 

The results obtained so far provide some tentative answers to our initial questions. What is the impact of P2P communication on information quality, when agents are exposed to central media information flooding? It depends on how pervasive the broadcasting is: peer-to-peer communication can contrast the impact of the complementary system until when no more than 60\% of the population is reached by the broadcasted messages. How much experts must be accessed through the P2P system for contrasting informational cheating? It depends on agents' confidence in their own opinions: when this is too high or too low, a larger number of experts (30\% or more) is needed to contrast informational cheating. When confidence is neither too high nor too low, even a small percentage (around 20\% ) is enough to obtain the same results.

Is peer-to-peer communication able to inhibit the information flooding exercised by the central media, inhibiting possible information cheating and containing the corruption of information? Our simulation provides a preliminary positive answer to this question. A white zone, in which there are no televiewers nor wise agents and in which agents can access information only through their neighbors, prevents opinions from converging on the values transmitted through the central media. Peer-to-peer communication matters in reducing aggregation of opinions.

As several recent research works have outlined out the importance of the dynamic aspects of social interactions \cite{Shao09,Lorenz2008,brunetti2010,AQ2010a,CFQS2010a,ACFQS2010a}, in future studies we are interested to characterise the opinions' evolution and how their behavior is affected by the dynamic nature of contacts.
In addition, we aim at implementing a more plausible model of opinions, taking into account other dimensions beside confidence, in particular the perceived correspondence between own and others' opinions and how these dimensions interact in the dynamics of beliefs.

\section{Acknowledgements}
This work was supported by the European Community under the FP6
programme (eRep project CIT5-028575). A particular thanks to Federica
Mattei, Mario Paolucci, Federico Cecconi, Stefano Picascia, Geronimo Stilton and the Hypnotoad.

\bibliographystyle{plain}
\bibliography{biblio}

\begin{thebibliography}{10}

\bibitem{ACFQS2010a}
Frederic Amblard, Arnaud Casteigts, Paola Flocchini, Walter Quattrociocchi, and
  Nicola Santoro.
\newblock Time varying graphs and temporal metrics for dynamic networks
  analysis.
\newblock {\em Technical Report University of Carleton, Canada}, 2010.

\bibitem{Asch55}
S.~E. Asch.
\newblock {Opinions and social pressure}.
\newblock {\em Scientific {A}merican}, 193:31--35, 1955.

\bibitem{brunetti2010}
S.~Brunetti, E.~Lodi, and W.~Quattrociocchi.
\newblock Multicolored dynamos on toroidal meshes.
\newblock {\em CoRR}, abs/1012.4404, 2010.

\bibitem{CFQS2010a}
A.~Casteigts, P.~Flocchini, W.~Quattrociocchi, and N.~Santoro.
\newblock {Time-Varying Graphs and Dynamic Networks}.
\newblock {\em Arxiv preprint arXiv:1012.0009}, 2010.

\bibitem{Castellano2007}
Claudio Castellano, Santo Fortunato, and Vittorio Loreto.
\newblock {Statistical physics of social dynamics}.
\newblock {\em Reviews of Modern Physics}, 81(2):591+, June 2009.

\bibitem{cavalli-sforza86}
L.~Cavalli-Sforza.
\newblock Cultural evolution.
\newblock {\em Amer. Zool. (1986) 26 (3): 845-856}, 1986.

\bibitem{amblard01}
G.~Deffuant, D.~Neau, F.~Amblard, and Gerard Weisbuch.
\newblock Mixing beliefs among interacting agents.
\newblock {\em Advances in Complex Systems}, 3:87--98, 2001.

\bibitem{Deighton95}
J.~Deighton.
\newblock {Marketing and Seduction: Building Exchange Relationships by Managing
  Social Consensus}.
\newblock {\em Journal of Consumer Research}, 21(4):660--676, March 1995.

\bibitem{festinger50}
L.~Festinger, S.~Schachter, and K.~Back.
\newblock {\em Social Pressures in Informal Groups: A Study of Human Factors in
  Housing}.
\newblock Harper, New York, NY, USA, 1950.

\bibitem{Harris04}
Richard~Jackson. Harris.
\newblock {\em A cognitive psychology of mass communication / by Richard
  Jackson Harris}.
\newblock L. Erlbaum Associates, Hillsdale, NJ :, 1989.

\bibitem{Hoadley99}
Christopher~M. Hoadley and Noel Enyedy.
\newblock Between information and communication: Middle spaces in computer
  media for learning, 1999.

\bibitem{Hu09}
Hai-Bo Hu and Xiao-Fan Wang.
\newblock {Discrete opinion dynamics on networks based on social influence}.
\newblock {\em Journal of Physics A: Mathematical and Theoretical},
  42(22):225005+, June 2009.

\bibitem{Balke08}
S.~Konig, T.~Balke, W.~Quattrociocchi, M.~Paolucci, and T.~Eymann.
\newblock On the effects of reputation in the internet of services.
\newblock In {\em ICORE 2009}. Gargonza, Italy, 2009.

\bibitem{Lorenz07}
Jan Lorenz.
\newblock {Continuous opinion dynamics of multidimensional allocation problems
  under bounded confidence: More dimensions lead to better chances for
  consensus}.
\newblock Aug 2007.

\bibitem{Lorenz2008}
Jan Lorenz and Dirk~A. Lorenz.
\newblock On conditions for convergence to consensus.
\newblock 2008.

\bibitem{Mason07}
Winter~A. Mason, Frederica~R. Conrey, and Eliot~R. Smith.
\newblock {Situating Social Influence Processes: Dynamic, Multidirectional
  Flows of Influence Within Social Networks}.
\newblock {\em Personality and Social Psychology Review}, 11(3):279--300,
  August 2007.

\bibitem{Maxwell}
Maxwell~E. Mccombs and Donald~L. Shaw.
\newblock {The Agenda-Setting Function of Mass Media}.
\newblock {\em The Public Opinion Quarterly}, 36(2):176--187, 1972.

\bibitem{niguez09opinion}
Gerardo~I. Niguez, J\'{a}nos Kert\'{e}sz, Kimmo~K. Kaski, and R.~A. Barrio.
\newblock {Opinion and community formation in coevolving networks}.
\newblock {\em Physical Review E}, 80(6):066119+, Dec 2009.

\bibitem{nowak90}
Andrzej Nowak, Jacek Szamrej, and Bibb Latan.
\newblock {From Private Attitude to Public Opinion: A Dynamic Theory of Social
  Impact}.
\newblock {\em Psychological Review}, 97:362--376, 1990.

\bibitem{Latane90}
Andrzej Nowak, Jacek Szamrej, and Bibb Latane.
\newblock {From Private Attitude to Public Opinion: A Dynamic Theory of Social
  Impact}.
\newblock {\em Psychological Review}, 97:362--376, 1990.

\bibitem{eRep}
M.~Paolucci, T.~Eymann, W.~Jager, J.~Sabater-Mir, R.~Conte, S.~Marmo,
  S.~Picascia, W.~Quattrociocchi, T.~Balke, S.~Koenig, T.~Broekhuizen,
  D.~Trampe, M.~Tuk, I.~Brito, I.~Pinyol, and D.~Villatoro.
\newblock {\em Social Knowledge for e-Governance: Theory and Technology of
  Reputation}.
\newblock Roma: ISTC-CNR, 2009.

\bibitem{AQ2010a}
W.~Quattrociocchi and F.~Amblard.
\newblock Selection in scientific networks.
\newblock Dec.

\bibitem{quattrociocchi2010d}
W.~Quattrociocchi, R.Conte, and E.Lodi.
\newblock Turning information into knowledge: The role of peer to peer
  communication.
\newblock {\em WCCS - World Congress on Social Simulation - Kessel, Germany},
  2010.

\bibitem{quattrociocchi2008}
Walter Quattrociocchi, Mario Paolucci, and Rosaria Conte.
\newblock Dealing with uncertainty :simulating reputation in an ideal
  marketplace.
\newblock In {\em AAMAS 08 Trust Workshop Cascais Portugal}, 2008.

\bibitem{quattrociocchi2009b}
Walter Quattrociocchi, Mario Paolucci, and Rosaria Conte.
\newblock Image and reputation coping differently with massive informational
  cheating.
\newblock In {\em WSKS (1)}, pages 574--583, 2009.

\bibitem{QuattrociocchiPC09}
Walter Quattrociocchi, Mario Paolucci, and Rosaria Conte.
\newblock On the effects of informational cheating on social evaluations: image
  and reputation through gossip.
\newblock {\em IJKL}, 5(5/6):457--471, 2009.

\bibitem{quattrociocchi2009a}
Walter Quattrociocchi, Mario Paolucci, and Rosaria Conte.
\newblock On the effects of informational cheating on social evaluations: image
  and reputation through gossip.
\newblock {\em IJKL}, 5(5/6):457--471, 2009.

\bibitem{quattrociocchi2008b}
Walter Quattrociocchi, Mario Paolucci, and Rosaria Conte.
\newblock {\em LNAI special issue on "Trust in Agent Societies"}, chapter
  Reputation and Uncertainty Reduction: Simulating Partner Selection.
\newblock Springer, pres.

\bibitem{Shao09}
Zhen Shao and Haijun Zhou.
\newblock Dynamics-driven evolution to structural heterogeneity in complex
  networks.
\newblock {\em Physica A: Statistical Mechanics and its Applications},
  388(4):523 -- 528, 2009.

\bibitem{Sherif88}
Muzafer Sherif.
\newblock {\em {The Robbers Cave Experiment: Intergroup Conflict and
  Cooperation}}.
\newblock Wesleyan, 1st wesleyan ed edition, February 1988.

\bibitem{Staufer2007}
D.~Stauffer.
\newblock {Opinion Dynamics and Sociophysics}.
\newblock May 2007.

\end{thebibliography}

\end{document}